\documentclass[12pt,a4paper]{article}
\usepackage[english]{babel}
\usepackage{latexsym,epsfig}
\usepackage{amsmath}

0

\begin{document}
\baselineskip 6,5mm

\def\II{\relax{\rm 1\kern-.35em1}}
\def\IP{\relax{\rm I\kern-.18em P}}
\def\muh{\hat \mu}
\def\nuh{\hat \nu}
\def\rhoh{\hat \rho}
\def\lambdah{\hat \lambda}
\def\sigmah{\hat \sigma}
\def\tauh{\hat \tau}

\begin{flushright}
hep-th/0106055 \\
IC/2001/51
\end{flushright}

\begin{center}

{}~\vfill

{ {\LARGE {Branes Wrapped on Coassociative Cycles}}}

\vspace{20 mm}

{\bf{\large{Rafael Hern\'{a}ndez}}}$^{\dag}$

\vspace{8 mm}

{\em The Abdus Salam International Center for Theoretical Physics \\
Strada Costiera, 11. $34014$ Trieste, Italy}
\vspace{16 mm}

\end{center}


\begin{center}
{\bf Abstract}
\end{center}
  
\vspace{2 mm}
  
We obtain a supergravity solution arising when D6-branes are wrapped on 
coassociative four-cycles of constant curvature in seven manifolds of $G_2$ holonomy. 
The solutions preserve two supercharges and thus represent supergravity duals of 
three dimensional Yang-Mills with $N=1$ supersymmetry. When uplifted to eleven 
dimensions our solution describes M-theory on the background of an eight manifold 
with Spin(7) holonomy.

\vspace{42 mm}

{\footnotesize \dag}\hspace{1 mm}{\footnotesize{\ttfamily e-mail address: rafa@ictp.trieste.it}}

\newpage


An interesting possibility to construct gravity duals of field theories with low supersymmetry 
is that provided by branes wrapped on supersymmetric cycles. As cycles will not in general 
have covariantly constant spinors, supersymmetry will only be preserved after an 
identification of the spin connection on the cycle with some external $R$-symmetry gauge fields; 
this identification defines a topologically twisted supersymmetric field theory \cite{BSV} 
(a detailed classification of different twists can be found in \cite{BT}). The 
way the cycle is embedded in a higher dimensional manifold determines the amount of 
preserved supersymmetry. When the number of branes is large, the uplifts to ten or eleven 
dimensions of the solutions, found in an adequate gauged supergravity, represent a gravity dual 
description of field theories with reduced supersymmetry. The case originally considered 
by Maldacena and N\'u\~nez \cite{MN1,MN2} was that of fivebranes and D3-branes wrapped 
on holomorphic curves, and has been applied in a series of related works to different 
dimension branes wrapped on diverse supersymmetric cycles \cite{AGK}-\cite{GKPW}. 
  
In \cite{EN}, Edelstein and N\'u\~nez studied a configuration of D6-branes wrapping holomorphic 
two-cycles and special Lagrangian three-cycles. When the size of the cycles is taken to zero, 
their solutions represent respectively a supergravity description of the infrared dynamics 
of five dimensional $N=2$ supersymmetric Yang-Mills, or four dimensional Yang-Mills with 
$N=1$ supersymmetry. D6-branes wrapping an $S^3$ in $T^* S^3$ had previously been proposed 
to be dual through a conifold transition to a type IIA geometry with the D6-branes replaced 
by RR fluxes on the blown up $S^2$ \cite{Vafa}. However, a better understanding of this 
duality came in terms of M-theory on a seven manifold of $G_2$ holonomy \cite{Acharya}, 
where it corresponds to an $S^3$ flop transition \cite{AMV} (see also \cite{Rolling}-\cite{Dasgupta} 
for further recent developments). These results were extended by Gomis in 
\cite{Gomis}, where it was argued how compactifications of M-theory on manifolds with 
reduced holonomy arise as the local eleven dimensional description of D6-branes wrapped 
on supersymmetric cycles in manifolds of lower dimension and with a different holonomy group. 
The authors of \cite{EN} explicitly reproduced the geometry of a manifold with $G_2$ holonomy and 
of the small resolution of the conifold when uplifting to eleven dimensions the solutions they found in 
eight dimensional maximal gauged supergravity, which is the natural arena to perform twisting 
for D6-branes.
  
The purpose of this letter is to use the approach of \cite{EN} to study one of the lifts 
considered in \cite{Gomis}, namely D6-branes wrapped on a coassociative four-cycle in a seven 
manifold of $G_2$ holonomy, which were shown to lift to M-theory on an eight manifold with 
Spin(7) holonomy group. Coassociative four-cycles are supersymmetric cycles preserving $1/16$ 
supersymmetry. Therefore, a collection of D6-branes wrapped on a coassocitive cycle will lead 
to a three dimensional gauge theory with $N=1$ supersymmetry.
  
In this letter we will construct a supergravity solution corresponding to D6-branes wrapped 
on a coassociative four-cycle, which represents a supergravity dual of a three dimensional 
gauge theory with two supercharges. The lift to eleven dimensions of this solution, following 
an argument identical to that in \cite{EN}, is then shown to correspond to an M-theory 
background which is a direct product of three dimensional Minkowski space and a manifold 
with Spin(7) holonomy. In order to do so, we will first shortly review maximal gauged 
supergravity in eight dimensions.
  
Maximal gauged supergravity in eight dimensions was constructed by Salam and Sezgin \cite{Salam}
through Scherk-Schwarz compactification \cite{Scherk} of eleven dimensional supergravity on 
an $SU(2)$ group manifold. The field theory content in the gravity sector of the theory 
\footnote{The fields arising from reduction of the eleven dimensional three-form are a scalar 
$B$, three vector fields $B_1^{i}$, three two-forms $B_2^{i}$ and a three-form $B_3$. However, 
we will only consider pure gravitational solutions of the eleven dimensional theory, so that 
all $B$ fields can be set to zero.} consists of the metric $g_{\mu \nu}$, a dilaton 
$\Phi$, five scalars given by a unimodular $3 \times 3$ matrix $L_{\alpha}^{i}$ in the coset 
$SL(3, {\bf R})/SO(3)$ and an $SU(2)$ gauge potential $A_{\mu}^{i}$, besides from the pseudo Majorana 
spinors $\psi_{\mu}$ and $\chi_i$ on the fermion side.
  
The Lagrangian for the bosonic fields is given, in $\kappa=1$ units, by 
\begin{equation}
e^{-1} {\cal L} = \frac {1}{4} R - \frac {1}{4} e^{2 \Phi} F_{\mu \nu}^{i} F^{\mu \nu \; i} - 
\frac {1}{4} P_{\mu \; ij} P^{\mu \; ij} - \frac {1}{2} (\partial_{\mu} \Phi)^2 - 
\frac {g^2}{16} e^{-2 \Phi} ( T_{ij} T^{ij} - \frac {1}{2} T^2),
\label{1}
\end{equation}
with $e$ the determinant of the achtbein $e^{a}_{\mu}$ and $F_{\mu \nu}^{i}$ the Yang-Mills 
field strenght. The Cartan decomposition of the $SL(3, {\bf R})/SO(3)$ coset defines the 
symmetric and traceless quantity $P_{\mu \, ij}$, as well as its antisymmetric counterpart, 
$Q_{\mu \, ij}$,
\begin{equation}
P_{\mu \, ij} + Q_{\mu \, ij} \equiv L_i^{\alpha} ( \partial_{\mu} \delta_{\alpha}^{\, \beta} 
- g \, \epsilon_{\alpha \beta \gamma} A^{\gamma}_{\mu}) L_{\beta \; j},
\label{2}
\end{equation}
which depends on the scalars parameterizing the coset and on the $SU(2)$ gauge fields. The 
potential energy associated to the scalar fields is given by the $T$-tensor
\begin{equation}
T^{ij} \equiv L^{i}_{\alpha} L^{j}_{\beta} \delta^{\alpha \beta},
\label{3}
\end{equation}
and $T \equiv T_{ij} \delta^{ij}$. Note that curved directions are labelled by greek indices, while 
flat ones are labelled by latin, and that $\mu, a = 0,1, \ldots, 7$ are spacetime coordinates, 
while $\alpha, i = 8,9,10$ are in the group manifold.
  
Bosonic solutions to the equations of motion, 
\begin{eqnarray}
R_{\mu \nu} \! \! & = & \! \! P_{\mu \, ij} P_{\nu}^{\, ij} + 2 \partial_{\mu} \Phi \partial_{\nu} 
\Phi + 2 e^{2 \Phi} F_{\mu \gamma}^{i} F_{\nu}^{\, \gamma i} - \frac {1}{3} g_{\mu \nu} 
\nabla^2 \Phi, \nonumber \\
\nabla_{\mu}(e^{2 \Phi}F^{\mu \nu i}) \! \! & = & \! \! -e^{2 \Phi} P_{\mu}^{ij} F_j^{\mu \nu} - 
g \, g^{\nu \gamma} \epsilon^{ijk} P_{\gamma \, jl} T_k^{\, l}, \nonumber \\
\nabla_{\mu} P^{\mu \, ij} \! \! & = & \! \! - \frac {2}{3} \delta^{ij} \nabla^2 \Phi + 
e^{2 \Phi} F_{\mu \nu}^{i} F^{\mu \nu \, j} + \frac {g^2}{2} e^{-2 \Phi} \Theta^{ij},
\end{eqnarray}
with $\Theta^{ij}$ a combination of the T-tensor
\begin{equation}
\Theta^{ij} \equiv T^i_k T^{jk} - \frac {1}{2} T T^{ij} - \frac {1}{2} \delta^{ij} 
(T_{kl} T^{kl} - \frac {1}{2} T^2),
\end{equation}
preserve supersymmetry if the supersymmetry 
variations for the fermions vanish,
\begin{eqnarray}
\delta \psi_{\gamma} \! \! & = & \! \! {\cal D}_{\gamma} \epsilon + \frac {1}{24} e^{\Phi} F_{\mu \nu}^{i} 
\Gamma_i ( \Gamma_{\gamma}^{\: \: \mu \nu} - 10 \delta_{\gamma}^{\, \mu} \Gamma^{\nu}) \epsilon 
- \frac {g}{288} e^{- \Phi} \epsilon_{ijk} \Gamma^{ijk} \Gamma_{\gamma} T \epsilon = 0, \nonumber \\
\delta \chi_i \! \! & = & \! \! \frac {1}{2} (P_{\mu \; ij} + \frac {2}{3} \delta_{ij} \partial_{\mu} 
\Phi) \Gamma^{j} \Gamma^{\mu} \epsilon - \frac {1}{4} e^{\Phi} F_{\mu \nu \; i} 
\Gamma^{\mu \nu} \epsilon - \frac {g}{8} e^{-\Phi} (T_{ij} - \frac {1}{2} \delta_{ij} T) 
\epsilon^{jkl} \Gamma_{kl} \epsilon = 0,
\label{4}
\end{eqnarray}
where the covariant derivative is
\begin{equation}
{\cal D}_{\mu} \epsilon = \partial_{\mu} \epsilon + \frac {1}{4} \omega_{\mu}^{ab} \Gamma_{ab} 
\epsilon + \frac {1}{4} Q_{\mu \; ij} \Gamma^{ij} \epsilon.
\label{5}
\end{equation}
  
A convenient representation for the Clifford algebra will be
\begin{equation}
\Gamma^{a} = \gamma^{a} \times \II, \: \: \: \: \: \: \: \: \Gamma^{i} = \gamma_9 \times \sigma^{i},
\label{6}
\end{equation}
where $\gamma_9 = i \gamma^0 \gamma^1 \ldots \gamma^7$, so that $\gamma_9^2 = \II$, and $\sigma^{i}$ 
are Pauli matrices. Furthermore, it will prove useful to introduce 
$\Gamma_9 \equiv \frac {1}{6i} \epsilon_{ijk} \Gamma^{ijk} = \gamma_9 \times \II$.
  
In this letter we are going to consider D6-branes wrapped on a coassociative four-cycle $S^4$ 
in a seven manifold of $G_2$ holonomy. The spin connection for the coassociative four-cycle is 
$SO(4)$. When we wrap the D6-branes on the four-cycle the $SO(1,6) \times SO(3)_R$ 
symmetry group of the unwrapped branes splits as $SO(1,2) \times SO(4) \times SO(3)_R$. The 
twisting is performed by identifying the structure group of the normal bundle, $SO(3)_R$, 
with $SU(2)_L$ in $SO(4) \simeq SU(2)_L \times SU(2)_R$. This leads to a pure gauge theory in three 
dimensions with two supercharges. There are no scalars \cite{Gomis} because the bundle of 
anti self-dual two-forms is trivial (which amounts to taking the four-sphere rigid as a coassociative 
submanifold \cite{McLean}).
  
In order to describe the deformation on the worldvolume of the D6-brane we will choose the 
metric ansatz 
\begin{equation}
ds^2 = e^{2f} dx_{1,2}^2 + dr^2 + e^{2h} ds_4^2,
\label{7}
\end{equation}
where the four-sphere metric will be taken as de Sitter's metric on $S^4$,
\begin{equation}
ds_4^2 = \frac {a^4}{(a^2 + \xi^2)^2} ( d \xi^2 + \xi^2 (\omega_1^2 + \omega_2^2 + \omega_3^2)),
\label{8}
\end{equation}
with $\omega_i$ the left-invariant one-forms on $SU(2)$ as a group manifold. The parameter $a$ is 
the diameter of the four-sphere, and will be later on identified with the instanton size. From 
the structure equations, the $O(4)$ connections $\omega_{ab}$ of $S^4$ can be easily shown to be
\begin{equation}
\omega_{4 \, i+4} = \frac {a^2 -\xi^2}{a^2 + \xi^2} \omega_i, \: \: \: \: \omega_{67} = \omega_1, 
\: \: \omega_{75} = \omega_2, \: \: \omega_{56} = \omega_3.
\label{9}
\end{equation}
  
The twisting amounts to an identification of the spin connection with the $R$-symmetry. In this case, 
it is possible to get rid of the scalars $L^{i}_{\alpha}$,
\begin{equation}
L^{i}_{\alpha} = \delta_{\alpha}^{i},
\label{10}
\end{equation}
so that
\begin{equation}
P_{ij} = 0, \: \: \: \: Q_{ij} = - g \epsilon_{ijk} A^k.
\label{11}
\end{equation}
Thus, the twisting is performed by turning on an $SU(2)$ gauge field obtained by identifying the self-dual 
combinations of the spin connection on $S^4$ with $Q_{ij}$, 
$A^{1} = - \frac {1}{g} (- \omega_{45} - \omega_{67})$ (+ cyclic), where 
$-\omega_{45}-\omega_{67}$ (+ cyclic) are self-dual combinations of the spin 
connection on $S^4$ \footnote{This construction is simply related to the fact that the instanton with 
unit second Chern number is the Hopf fibration of $S^7$.}. The gauge field is then that 
for the charge one $SU(2)$ instanton on $S^4$,
\begin{equation}
A = \frac {1}{g} \frac {a^2}{a^2 + \xi^2} i \, \omega_i \sigma^{i}.
\label{13}
\end{equation}
Imposing the projections on a coassociative cycle \cite{GLW}
\begin{equation}
\gamma_{45}^- \epsilon = \Gamma^{23} \epsilon, \: \: \gamma_{46}^- \epsilon =
\Gamma^{31} \epsilon, \: \: \gamma_{47}^- \epsilon = \Gamma^{12} \epsilon,
\label{14}
\end{equation}
and 
\begin{equation}
\gamma_{ab}^+ \epsilon = 0,
\label{15}
\end{equation}
where the minus and plus signs refer to anti self-dual and self-dual parts, respectively, and 
$a,b = 4,5,6,7$, as well as
\begin{equation}
\gamma_r \epsilon = - i \gamma_9 \epsilon,
\label{16}
\end{equation}
together with (\ref{13}) for the gauge field, leads the BPS equations to
\begin{eqnarray}
f' \! \! & = & \! \! \frac {\Phi'}{3} \, = \, - \frac {1}{g a^2} e^{\Phi-2h} + 
\frac {g}{8} e^{-\Phi}, \nonumber \\
h' \! \! & = & \! \! \frac {2}{g a^2} e^{\Phi-2h} + \frac {g}{8} e^{-\Phi}.
\label{17}
\end{eqnarray}
  
After the change of variables
\begin{equation}
r(\rho) = \frac {(ga^3)^{1/2}}{2} \sqrt{\frac {3}{5}} \Big[ \frac {2}{3} \rho^{2/3}_{\: \quad 2} F_1 \big[ 
- \frac {9}{20}, \frac {1}{4}, \frac {11}{20}; \frac {l^{10/3}}{\rho^{10/3}} \big] + 
3 l^{3/2} \frac {\Gamma\left(-\frac {9}{20} \right) \Gamma\left( \frac {3}{4} \right)}{10 
\Gamma\left(\frac {3}{10} \right)} \Big],
\label{18}
\end{equation}
a solution to the BPS equations can be shown to be 
\begin{equation}
e^{2 \Phi} = (ga)^3 \left( \frac {3}{20} \right)^3 \! \rho^3 \left( 1 - \frac {l^{10/3}}{\rho^{10/3}} 
\right)^{3/2}, \: \: \: \: e^{2h} = ga\frac {27}{400} \rho^3 \left(1 - 
\frac {l^{10/3}}{\rho^{10/3}} 
\right)^{1/2},
\label{19}
\end{equation}
where $l^{10/3}$ is an integration constant.
  
The lift to eleven dimensions of this solution, using the elfbein in \cite{Salam}, leads to 
\footnote{It is immediate to check that the simpler solution $e^{2 \Phi} = g^2a^2/20 \, r^2$, 
$e^{2 h} = r^2$ to the system of equations (\ref{17}), when lifted to eleven dimensions, reproduces a 
solution which is three dimensional Minkowski space times a metric whose level surfaces 
$r=\hbox {constant}$ tend to the homogeneus squashed Einstein metric on the seven-sphere, 
as already noted in \cite{GPP} concerning the large $\rho$ limit of (\ref{20}).}
\begin{equation}
ds_{11}^2 = dx_{1,2}^2 + \frac {d \rho^2}{\left( 1 - \frac {l^{10/3}}{\rho^{10/3}} 
\right)} + \frac {9}{100} \rho^2 \left( 1 - \frac {l^{10/3}}{\rho^{10/3}} 
\right) (\tilde{\omega}_{i} - A^{i})^2 + \frac {9}{20} \rho^2 ds_4^2,
\label{20}
\end{equation}
which is the metric of a Spin(7) holonomy manifold \cite{BS,GPP} \footnote{See also 
\cite{Floratos}-\cite{Konishi} for different approaches to this metric.}, 
with the topology of an ${\bf R}^4$ bundle over $S^4$.
  
We have thus been able to reproduce, by studying the M-theory description of a 
configuration of D6-branes wrapped on a coassociative submanifold, the 
metric constructed in \cite{BS,GPP} for an eight manifold with Spin(7) holonomy. This 
was one of the lifts already proposed in \cite{Gomis}, where it was shown how 
there is an M-theory realization involving Spin(7) holonomy of the strong coupling 
description of D6-branes wrapped on a coassociative cycle. 
  
Recently new explicit metrics on complete non compact Riemann eight manifolds with Spin(7) 
holonomy have been constructed \cite{Cveticspin7}. As a difference with the previously 
known metric of \cite{BS,GPP}, the ones found in \cite{Cveticspin7} exhibit an asymptotically locally 
conical behavior. It would be interesting to understand this feature and to reproduce the metrics 
using a lift to eleven dimensions of some eight dimensional supergravity solution.


\vspace{8 mm}

{\bf Acknowledgements}

It is a pleasure to thank M. Blau, C. G\'omez, K. Narain and S. Randjbar-Daemi 
for useful discussions. This research is partly supported by the EC contract no. HPRN-CT-2000-00148.

\newpage


\end{document}